\documentclass[12pt]{article}
\usepackage{epsfig}
\usepackage{color}
\usepackage{amssymb,amsmath}
\usepackage{graphicx}
\usepackage{here}
\newcommand{\bea}{\begin{eqnarray}}
\newcommand{\eea}{\end{eqnarray}}
\newcommand{\nn}{\nonumber}

 \makeatletter
\def\alt{\mathrel{\mathpalette\gl@align<}}
\def\agt{\mathrel{\mathpalette\gl@align>}}
\def\gl@align#1#2{\lower.6ex\vbox{\baselineskip\z@skip\lineskip\z@
\ialign{$\m@th#1\hfil##\hfil$\crcr#2\crcr\sim\crcr}}} \makeatother

\begin{document}
\begin{flushright}
\end{flushright}
\vspace*{1.0cm}

\begin{center}
\baselineskip 20pt 
{\Large\bf 
Muon and Electron $g-2$ \\
 and the Origin of Fermion Mass Hierarchy
}
\vspace{1cm}

{\large 
Naoyuki Haba, \ Yasuhiro Shimizu \ and \ Toshifumi Yamada
} \vspace{.5cm}

{\baselineskip 20pt \it
Graduate School of Science and Engineering, Shimane University, Matsue 690-8504, Japan
}

\vspace{.5cm}

\vspace{1.5cm} {\bf Abstract} \end{center}

We present a model that  gives a natural explanation to the charged lepton mass hierarchy
and study the contributions to the electron and the muon $g-2$.
In the model, we introduce lepton-flavor-dependent $U(1)_F$ symmetry and three additional Higgs doublets with $U(1)_F$ charges,
 to realize that each generation of charged leptons couples to one of the three additional Higgs doublets.
The $U(1)_F$ symmetry is softly broken by $+1$ charges,
 and the smallness of the soft breaking naturally gives rise to the hierarchy of the Higgs VEVs, which then accounts for the charged lepton mass hierarchy.
 Since electron and muon couple to different scalar particles, each scalar contributes to
 the electron and the muon $g-2$ differently. 
 We survey the space of parameters of the Higgs sector and find that there
are sets of parameters that explain the muon  $g-2$ discrepancy.
 On the other hand, we cannot find the parameter sets that can explain $g-2$ discrepancy
 within 2$\sigma$.
Here the $U(1)_F$ symmetry suppresses charged lepton flavor violation.

\thispagestyle{empty}

%\bigskip
\newpage

%\addtocounter{page}{-1}
\setcounter{footnote}{0}
%%%%%%%%%%%%%%%%%%%%%%%%%%
%\baselineskip 36pt
% Main body
%%%%%%%%%%%%%%%%%%%%%%%%%%
\baselineskip 18pt
%%%%%%%%%%%%%%%%%%%%%%%%%%
%

\section{Introduction}
The Standard Model (SM) explains almost all experimental data, but there are
several anomalies in low-energy observables.
Among the anomalies, the experimental and theoretical studies on the muon $g-2$ 
has been extensively carried out.
Comparing the experimental value to its SM prediction, 
the current discrepancy of the muon $g-2$ is reported as \cite{Davier:2017zfy,Keshavarzi:2018mgv,Bennett:2006fi,Roberts:2010cj}
\bea
\delta a_\mu &=& a_\mu^\mathrm{obs}-a_\mu^\mathrm{SM}=(27.4\pm 7.3)\times 10^{-10}.
\label{amu_exp}
\eea
Recently, the discrepancy of the electron $g-2$ is also reported as \cite{Hanneke:2008tm,Hanneke:2010au,Aoyama:2014sxa,Parker:2018vye}
\bea
\delta a_e &=& a_e^\mathrm{obs}-a_e^\mathrm{SM}=(-8.7\pm 3.6)\times 10^{-13}.
\label{ae_exp}
\eea
Although the electron $g-2$ discrepancy is less than 3$\sigma$, both discrepancies may be
signs of physics beyond the SM.

In the effective theoretical approach, the contribution to the lepton $g-2$  is described by the chirality-breaking operator, $\sim \overline{\ell_L}\sigma_{\mu\nu}\ell_R F^{\mu\nu}$.
If the chirality-breaking in a new physics model is proportional simply to the lepton mass,
as in Minimal Supersymmetric Standard Model (MSSM) with Minimal Flavor Violation (MFV),
the contribution to the lepton $g-2$ is proportional to its mass, and 
there is a simple relation between the electron and muon $g-2$
\bea
\frac{\delta a_e}{\delta a_\mu}\simeq\frac{m_e^2}{m_\mu^2}=2.3\times 10^{-5}.
\label{ratio}
\eea
This relation does not hold experimentally in Eq.~(\ref{amu_exp}) and Eq.~(\ref{ae_exp}), 
especially the sign is opposite.
Hence, any new physics model that explains both discrepancies must include flavor violation beyond MFV in the interactions of muon and electron.
Various sources of flavor violation beyond MFV have been considered to address the discrepancies
\cite{Lindner:2016bgg,Davoudiasl:2018fbb,Crivellin:2018qmi,Liu:2018xkx,Dutta:2018fge,Han:2018znu,Endo:2019bcj,Badziak:2019gaf,Abdullah:2019ofw}.

A stringent restriction~\cite{Crivellin:2018qmi} on the model building attempts comes from the absence of the $\mu\to e\gamma$ decay.
In the effective theory point of view, this indicates that two operators $\overline{\mu_L}\sigma_{\mu\nu}\mu_R F^{\mu\nu}$ and $\overline{e_L}\sigma_{\mu\nu}e_R F^{\mu\nu}$ appear in a way that breaks MFV, 
 but the operators $\overline{e_L}\sigma_{\mu\nu}\mu_R F^{\mu\nu}$ and $\overline{\mu_L}\sigma_{\mu\nu}e_R F^{\mu\nu}$ are forbidden.
The above situation is realized naturally by assuming a muon-specific and/or electron-specific $U(1)$ symmetry.
One possibility is that this $U(1)$ symmetry is the same as the accidental muon-number and electron-number symmetries of the SM, namely,
 the new physics sector respects the muon-number and electron-number symmetries so that 
 $\overline{\mu_L}\sigma_{\mu\nu}\mu_R F^{\mu\nu}$ and $\overline{e_L}\sigma_{\mu\nu}e_R F^{\mu\nu}$ are generated with arbitrary strengths,
 but $\overline{e_L}\sigma_{\mu\nu}\mu_R F^{\mu\nu}$ and $\overline{\mu_L}\sigma_{\mu\nu}e_R F^{\mu\nu}$ are not generated.
Another possibility is that only $\mu_R$ or $e_R$ is charged under a new (anomalous) $U(1)$ symmetry, and there exists a new Higgs field charged under it
 that couples exclusively to $\mu_R$ or $e_R$.

In this paper, we construct a model along the second possibility to explain the electron and muon $g-2$ discrepancies without invoking large charged lepton flavor violation.
We go one step further and connect the new $U(1)$ symmetry to the origin of the fermion mass hierarchy.

In our model (which we name `lepton-flavored Higgs model'), 
 we introduce a new $U(1)_F$ symmetry under which $\tau_R,\mu_R,e_R$ are charged by $-1,-2,-3$, respectively,
 and introduce three additional Higgs doublets with $U(1)_F$ charges $+1,+2,+3$.
Due to the $U(1)_F$ symmetry, each generation of charged leptons couples to one of the three Higgs doublets.
We assume that the $U(1)_F$ symmetry is softly broken with a small amount by $+1$ charges, which naturally generates a hierarchy among 
the vacuum expectation values (VEVs) of the Higgs doublets.
We consider that this hierarchy of VEVs accounts for the charged lepton mass hierarchy and that the charged lepton Yukawa couplings are all $O(1)$.
A notable feature of the above setup is that since the charged and heavy neutral scalars in the Higgs sector couple differently to electron and muon,
 there is little correlation between the new scalar contributions to the electron and muon $g-2$.

We survey the space of parameters of the Higgs sector and find that there
are sets of parameters that explain the muon $g-2$ discrepancy within the 1$\sigma$ region.
Unfortunately, we cannot find any parameter set that explains the electron  $g-2$ discrepancy
within the 2$\sigma$ region.

This paper is organized as follows.
In Section~2, we explain our lepton-flavored Higgs model. 
In Section~3, we conduct a numerical search of parameters of the Higgs sector.
Section~4 summarizes the paper.
\\

\section{Lepton-Flavored Higgs Model}

The lepton-flavored Higgs model includes four Higgs doublets, $H_1,\,H_2,\,H_3,\,H_0$, 
 in addition to the SM leptons.
The fields are charged under the SM $SU(3)_C\times SU(2)_L\times U(1)_Y$ gauge group and a new anomalous $U(1)_F$ symmetry
 as Table~\ref{fields}.
Note that since only the right-handed charged leptons have $U(1)_F$ charges, the model does not restrict the Weinberg operator for the tiny neutrino mass.
\begin{table}[H]
\begin{center}
  \caption{The fields and their charge assignments. $\alpha$ labels the three generations.}
  \begin{tabular}{|c||c|c|c|c|} \hline
    Field & $SU(3)_C$ & $SU(2)_L$ & $U(1)_Y$ & anomalous $U(1)_F$ \\ \hline
    $H_1$  & {\bf 1} & {\bf 2}          & $+1/2$      & $+3$ \\
    $H_2$  & {\bf 1} & {\bf 2}          & $+1/2$      & $+2$ \\ 
    $H_3$  & {\bf 1} & {\bf 2}          & $+1/2$      & $+1$ \\ 
    $H_0$       & {\bf 1} & {\bf 2}          & $+1/2$      & 0    \\ \hline
    $\ell_L^\alpha$ $(\alpha=1,2,3)$ & {\bf 1} & \bf{2}          & $-1/2$      & 0 \\ \hline
    $e_R$              & {\bf 1} & \bf{1}          & $-1$      & $-3$ \\
    $\mu_R$              & {\bf 1} & \bf{1}          & $-1$      & $-2$ \\
    $\tau_R$              & {\bf 1} & \bf{1}          & $-1$      & $-1$ \\ \hline
    \end{tabular}
  \label{fields} 
  \end{center}
\end{table}

The $U(1)_F$ symmetry is assumed to be softly broken by $+1$ charges.
The soft breaking by $+1$ charges
 can be realized by introducing a SM gauge-singlet scalar $S$ with $U(1)_F$ charge $1/2$ and demanding that renormalizable interactions
 preserve the $U(1)_F$.
Then one introduces a VEV of $S$ to break the $U(1)_F$.
The resulting Nambu-Goldstone boson gains mass from non-renormalizable terms that explicitly violate the $U(1)_F$.

The Yukawa couplings for the SM leptons, which respect the $U(1)_F$ symmetry, are given by
\bea
-{\cal L}_{\rm Yukawa}&=&y_{1}\ \bar{\ell}_L^1 H_1 e_R+y_{2}\ \bar{\ell}_L^2 H_2 \mu_R+y_{3}\ \bar{\ell}_L^3 H_3 \tau_R
+{\rm H.c.}
\label{yukawa}
\eea
The Higgs potential, where the $U(1)_F$ symmetry is softly broken by $+1$ charges, is given by
\bea
-{\cal L}_{\rm Higgs}&=&m_1^2\,H_1^\dagger H_1+m_2^2\,H_2^\dagger H_2+m_3^2\,H_3^\dagger H_3+m_0^2\,H_0^\dagger H_0
\nn\\
&-&\mu_{12}^2(H_1^\dagger H_2+H_2^\dagger H_1)-\mu_{23}^2(H_2^\dagger H_3+H_3^\dagger H_2)
-\mu_{03}^2(H_0^\dagger H_3+H_3^\dagger H_0)
\nn\\
&+&
\lambda_1(H_1^\dagger H_1)^2+\lambda_2(H_2^\dagger H_2)^2+\lambda_3(H_3^\dagger H_3)^2
+\lambda_0(H_0^\dagger H_0)^2
\nn\\
&+&\sum_{i,j=0,1,2,3;i>j}
\lambda_{ij}(H_i^\dagger H_i)(H_j^\dagger H_j)
+\sum_{i,j=0,1,2,3;i>j}
\rho_{ij}(H_i^\dagger H_j)(H_j^\dagger H_i)
\nn\\
&+&
\kappa_1
\left(
(H_0^\dagger H_2)(H_1^\dagger H_3)+(H_2^\dagger H_0)(H_3^\dagger H_1)
\right)
+
\kappa_2
\left(
(H_0^\dagger H_3)(H_1^\dagger H_2)+(H_3^\dagger H_0)(H_3^\dagger H_2)
\right).~~ 
\label{potential}
\eea
For simplicity, we assume that the Higgs potential is CP invariant and all the parameters are real.

Now we make a crucial assumption on the Higgs potential.
We assume
\bea
m_0^2<0, \ \ \ m_1^2>0, \ \ \ m_2^2>0,  \ \ \ m_3^2>0.
\eea
$H_0$ develops a VEV, $\langle H_0\rangle=v/\sqrt{2}$, which is estimated to be
\bea
v \ \simeq \ \sqrt{\frac{-m_0^2}{\lambda_0}}.
\eea
The VEV of $H_0$ induces a VEV of $H_3$ through the term $\mu_{03}^2(H_0^\dagger H_3+H_3^\dagger H_0)$.
The latter induces a VEV of $H_2$ through the term $\mu_{23}^2(H_2^\dagger H_3+H_3^\dagger H_2)$,
 which then induces a VEV of $H_1$ through the term $\mu_{12}^2(H_1^\dagger H_2+H_2^\dagger H_1)$.
Consequently, writing the VEVs as $\langle H_3\rangle=v_3/\sqrt{2}$, $\langle H_2\rangle=v_2/\sqrt{2}$, $\langle H_1\rangle=v_1/\sqrt{2}$, we get
\bea
v_3 \ \simeq \ \frac{\mu_{03}^2}{m_3^2}v,
\ \ \ \ \ \ 
v_2 \ \simeq \ \frac{\mu_{23}^2}{m_2^2}v_3,
\ \ \ \ \ \ 
v_1 \ \simeq \ \frac{\mu_{12}^2}{m_1^2}v_2.
\label{vev}
\eea
We arrange the masses such that ($m_t$ denotes the top quark mass)
\bea
\frac{\mu_{03}^2}{m_3^2}\sim\frac{m_{\tau}}{m_t}\ll1, \ \ \ \ \
\frac{\mu_{23}^2}{m_2^2}\sim\frac{m_\mu}{m_\tau}\ll1, \ \ \ \ \ \frac{\mu_{12}^2}{m_1^2}\sim\frac{m_e}{m_\mu}\ll1.
\label{assumption2}
\eea
The arrangement of Eq.~(\ref{assumption2}) is natural because
 $\mu_{03}^2,\mu_{23}^2,\mu_{12}^2$ break the $U(1)_F$ symmetry while $m_3^2,m_2^2,m_1^2$ preserve it.
It follows from Eq.~(\ref{assumption2}) that the lepton Yukawa couplings are all $O(1)$,
\bea
y_{1} \sim y_{2} \sim y_{3} \sim 1.
\eea
We thus naturally explain the hierarchy of the charged lepton masses in terms of the small soft breaking of the $U(1)_F$ symmetry.

One can implement a similar structure in the quark sector, where we have five more Higgs doublets whose VEVs are on the order of
$(m_b/m_t)v$, $(m_s/m_b)v$, $(m_d/m_s)v$, $(m_c/m_t)v$, $(m_u/m_c)v$, respectively,
 and which couple exclusively to the right-handed bottom, strange, down, charm and up quarks, respectively.
Also, $H_0$ couples only to the right-handed top quark.
The resulting model is basically the same as the progressive $U(1)$ model of Ref.~\cite{Ma:2016zod}.
\\

We move to phenomenological aspects of the model.
After electroweak symmetry breaking, there appear four CP-even scalar particles, three CP-odd scalar particles, and three charged scalar particles.
Since $v_1\ll v_2\ll v_3 \ll v$, we can make an approximation that
 each CP-odd scalar particle comes exclusively from $H_1$, $H_2$ or $H_3$, 
 each CP-even scalar particle comes exclusively from $H_1$, $H_2$, $H_3$ or $H_0$,
 and each charged scalar particle comes exclusively from $H_1$, $H_2$ or $H_3$.
Under the above approximation, the Yukawa couplings in Eq.~(\ref{yukawa}) are rewritten in terms of physical particles as
\bea
{\cal L}_{\rm Yukawa}&\simeq&
\frac{y_{1}v_1}{\sqrt{2}}\,\bar{e} e
+\frac{y_{2}v_2}{\sqrt{2}}\,\bar{\mu} \mu
+\frac{y_{3}v_3}{\sqrt{2}}\,\bar{\tau} \tau
+\frac{y_{1}}{\sqrt{2}}\frac{v_1}{v}\,h\,\bar{e} e
+\frac{y_{2}}{\sqrt{2}}\frac{v_2}{v}\,h\,\bar{\mu} \mu
+\frac{y_{3}}{\sqrt{2}}\frac{v_3}{v}\,h\,\bar{\tau} \tau
\nn\\
&+&\frac{y_{1}}{\sqrt{2}}\,H_1^0\,\bar{e} e
+\frac{y_{2}}{\sqrt{2}}\,H_2^0\,\bar{\mu} \mu
+\frac{y_{3}}{\sqrt{2}}\,H_3^0\,\bar{\tau} \tau
\nn\\
&+&\frac{y_{1}}{\sqrt{2}}\,A_1\,\bar{e}i\gamma_5 e
+\frac{y_{2}}{\sqrt{2}}\,A_2\,\bar{\mu}i\gamma_5 \mu
+\frac{y_{3}}{\sqrt{2}}\,A_3\,\bar{\tau}i\gamma_5 \tau
\nn\\
&+&y_{1}\,H_1^+\,\bar{\nu}_e e_R
+y_{2}\,H_2^+\,\bar{\nu}_\mu \mu_R
+y_{3}\,H_3^+\,\bar{\nu}_\tau \tau_R
+{\rm H.c.}
\eea
 where $A_1,A_2,A_3$ denote CP-odd scalar particles, $h,H_1^0,H_2^0,H_3^0$ CP-even scalar particles, and $H_1^+,H_2^+,H_3^+$
 charged scalar particles.
$h$ has SM-like Yukawa couplings and can be identified with the observed 125~GeV scalar particle.

We concentrate on the contribution of $H_1^0,A_1,H_1^+$ to the electron $g-2$ and that of $H_2^0,A_2,H_2^+$ to the muon $g-2$.
They are given by~\cite{Dedes:2001nx}
\bea
\delta a_e &=& \frac{1}{16\pi^2}y_{1}^2\left\{
\frac{m_e^2}{m_{H_1^0}^2}\left(\log\frac{m_{H_1^0}^2}{m_e^2}-\frac{7}{6}\right)-
\frac{m_e^2}{m_{A_1}^2}\left(\log\frac{m_{A_1}^2}{m_e^2}-\frac{11}{6}\right)
-\frac{m_e^2}{6m_{H_1^+}^2}\right\},
\label{ae}
\\
\delta a_\mu &=& \frac{1}{16\pi^2}y_{2}^2\left\{
\frac{m_\mu^2}{m_{H_2^0}^2}\left(\log\frac{m_{H_2^0}^2}{m_\mu^2}-\frac{7}{6}\right)-
\frac{m_\mu^2}{m_{A_2}^2}\left(\log\frac{m_{A_2}^2}{m_\mu^2}-\frac{11}{6}\right)
-\frac{m_\mu^2}{6m_{H_2^+}^2}\right\}.
\label{amu}
\eea
It is important to note that different sets of scalar masses enter the formulas for the electron and muon $g-2$.
This allows us to simultaneously explain the negative deviation of the electron $g-2$ and the positive deviation of the muon $g-2$.
We comment that the two-loop Barr-Zee diagrams are suppressed by the electron mass or muon mass and hence are negligible.

We derive the masses of scalar particles $H_1^0,A_1,H_1^+$, $H_2^0,A_2,H_2^+$ from the scalar potential Eq.~(\ref{potential}),
 and show that there exists a parameter region where the discrepancies of electron and muon $g-2$ can be explained,
 with $O(1)$ values for Yukawa couplings $y_1,y_2$ and without conflicting experimental bounds on the masses of CP-even, CP-odd and charged scalar particles.
Expanding the Higgs potential around the VEVs in Eq.~(\ref{vev}), the tadpole parameters $t_{H_i}$ for the CP-even scalars should vanish. 
\bea
t_{H_1} &\simeq& m_1^2v_1-\mu_{12}^2v_2+\frac{\lambda_{10}+\rho_{10}}{2}v^2 v_1=0,
\label{tadpole1}
\\
t_{H_2} &\simeq& m_2^2v_2-\mu_{23}^2v_3+\frac{\lambda_{20}+\rho_{20}}{2}v^2 v_2=0,
\label{tadpole2}
\\
t_{H_3} &\simeq& m_3^2v_3-\mu_{30}^2v_0+\frac{\lambda_{30}+\rho_{30}}{2}v^2 v_3=0.
\label{tadpole3}
\eea
Here the hierarchy of the VEVs,  $v_1\ll v_2 \ll v_3 \ll v$, is assumed.
The physical Higgs mass spectrum is given by
\bea
m^2_{H_i^0}=m^2_{A_i}&\simeq& m_i^2+\frac{\lambda_{i0}+\rho_{i0}}{2}v^2,
\label{ma}
\\
m^2_{H_i^+} &\simeq& m_i^2+\frac{\lambda_{i0}}{2}v^2.
\label{mch}
\eea
 Notice that the CP-even and the CP-odd scalars have different masses due to the quartic coupling constants $\rho_{i0}$. 
\\

\section{Numerical Results}
First, let us estimate the contributions of the new scalar particles to $\delta a_e$ and $\delta a_\mu$.
From Eq.~(\ref{ae}) and Eq.~(\ref{amu}), 
 \bea
 \delta a_e &=& 2.5 \times 10^{-13} \times  \left(\frac{y_1}{3}\right)^2 \left(\frac{100\, \mathrm{GeV}}{m_{H_1^0}}\right)^2\left( 4- \left(\frac{m_{H^0_1}}{m_{H^+_1}}\right)^2\right)~\mathrm{e\,cm},
 \\
  \delta a_\mu &=& 1.2 \times 10^{-9} \times  \left(\frac{y_2}{1}\right)^2 \left(\frac{100\, \mathrm{GeV}}{m_{H_2^0}}\right)^2\left( 4- \left(\frac{m_{H^0_2}}{m_{H^+_2}}\right)^2\right)~\mathrm{e\,cm},
 \eea
 where $m_{A_i}=m_{H^0_i}$ is assumed. 
 In order to explain the magnitude of the current discrepancies in Eq.~(\ref{ae_exp}) and Eq.~(\ref{amu_exp}), the Yukawa coupling constants must be O(1) and the new scalar masses must be O(100) GeV.
 Since the current deviation of the electron $g-2$ is negative, there is a condition that
 \bea
  m_{H^0_1}>2m_{H^+_1}. 
 \eea
 From Eq.~(\ref{ma}) and Eq.~(\ref{mch}), this inequality can be satisfied if $\rho_{10}$ is $O(1)$.
 On the other hand, the current deviation of the muon $g-2$ is positive, so that
  \bea
   m_{H^0_2}<2m_{H^+_2}. 
 \eea
This inequality can be satisfied if $\rho_{20}$ is sufficiently small.

\begin{figure}[htbp]
\begin{center}
\includegraphics[width=15cm]{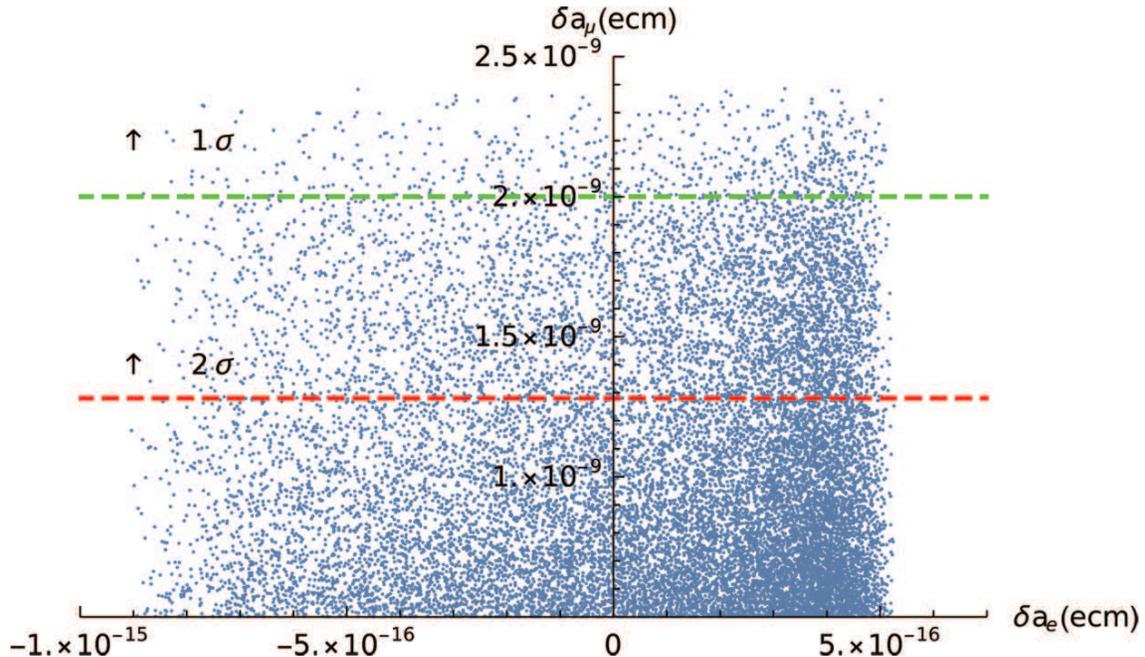}
\caption{Scatter plot of $\delta a_e$ and  $\delta a_\mu$. The Yukawa coupling constants are fixed $y_1=0.6$ and $y_2=3.3$ and other input parameters are explained in the text. The green and red lines correspond to the experimental 1$\sigma$ and  2$\sigma$ bounds, respectively.}
\label{fig}
\end{center}
\end{figure}

In the numerical analysis, we take the VEVs of the four Higgs doublets as follows.
\bea
v_3=\frac{m_\tau}{m_t}v,~v_2=\frac{m_\mu}{m_\tau}v_3,~v_1=\frac{m_e}{m_\mu}v_2,
\eea
with $v=246$ GeV. For the other Higgs sector parameters, we scan them randomly in the following regions.
\bea
0&<&m_1,~m_2,~m_3<50~\mathrm{GeV},
\\
0&<&\lambda_i,~\lambda_{ij},~\kappa_i<1,
\\
0&<&\rho_{ij}<6.
\eea
For each parameter set, we impose the vanishing tadpole conditions in Eqs.~(\ref{tadpole1})-(\ref{tadpole3}) and calculate the physical Higgs spectrum. 
We cannot find the LHC constraint on the lepton-specific Higgs. If the Yukawa couplings
are O(1), the charged Higgs mainly decay into the charged lepton and neutrino.
In this case, the charged Higgs search is almost the same as the left-handed slepton search.
The current LHC bound on the left-handed slepton is $\sim$ 300 GeV for the combination of selectron
and smuon \cite{Aad:2014vma,Khachatryan:2014qwa}. There is no separate mass bound
and we impose the following flavor dependent constraints.
\bea
m^+_{H_1}>200\, \mathrm{GeV},~~m^+_{H_{2,3}}>400\, \mathrm{GeV}.
\eea
If $y_1$ is large, the effective operator $\frac{y_1^2}{m^2_{H_1^+}} \overline{e}e\overline{\nu}\nu$ through the $H_1^+$ mediated diagram  is enhanced and
 there is a severe constraint on the effective operator
 from mono-photon search in LEP \cite{Fox:2011fx}.
 \bea
 m_{H^+_1}/y_1 >340\, \mathrm{GeV}.
 \eea
Fig.\ref{fig} shows the contributions of the new scalar particles to $\delta a_e$ and $\delta a_\mu$.
Here we fix the Yukawa coupling constants as $y_1=0.6$ and $y_2=3.3$.
We observe that there are parameter sets that give $\delta a_\mu$ in the 1$\sigma$ region. 
Although there are parameter sets where $\delta a_e$ is negative and reaches $\sim -9 \times 10^{-16}$ ecm,  we
cannot find any parameter set that gives $\delta a_e$ even in 
the 2$\sigma$ region.
 In these parameter sets, the charged scalar masses are around the current experimental
 bound  $\sim 200$ GeV and $\rho_{10}$ is $O(1)$.

\section{Summary}
We have studied the lepton-flavored Higgs model to explain the electron and the muon $g-2$ discrepancies simultaneously,
 and further give a natural explanation to the charged lepton mass hierarchy.
In the model, we introduce the $U(1)_F$ symmetry under which $e_R,\mu_R,\tau_R$ are charged with $-3,-2,-1$, respectively,
 and introduce three additional Higgs doublets with $U(1)_F$ charges $+3,+2,+1$,
 so that each generation of charged leptons couples to one of the three additional Higgs doublets due to the $U(1)_F$ symmetry.
We assume that the $U(1)_F$ symmetry is softly broken by $+1$ charges, and take the soft breaking to be sufficiently small
 that the hierarchy of the charged lepton masses are originated from the hierarchy of the Higgs VEVs while
 the Yukawa couplings are $O(1)$.
Note that it is natural to take the soft breaking small.
Since electron and muon couple to different scalar particles with O(1) Yukawa couplings,
 the electron and the muon $g-2$ discrepancies can be explained simultaneously. 
Specifically, the negative deviation of $a_e$ is explained when the almost-electron-specific charged scalar particle
is much lighter than the CP-even scalar particle. On the other hand, the positive deviation of
$a_\mu$ is explained when the mass of the almost-muon-specific charged scalar particle is similar to
that of the CP-even scalar particle. 
We have searched the space of the Higgs sector parameters and
found sets that give  $a_\mu$ in the 1$\sigma$ region. 
Although there are  parameter sets that give negative $\delta a_e$, we cannot find any parameter set 
that gives $a_e$ in the 2$\sigma$ region.

\section*{Acknowledgement}

This work is partially supported by Scientific Grants by the Ministry ofEducation, Culture,Sports, Science and Technology of Japan, Nos. 17K05415, 18H04590 and 19H051061 (NH), and No. 19K147101 (TY).

\end{document}